\documentclass[aps,prl,groupedaddress,superscriptaddress,twocolumn]{revtex4}

\usepackage{graphicx}
\usepackage{amssymb}

\begin{document}

\title{Hyperactivated resistance in TiN films on the insulating side\\
of the disorder-driven superconductor-insulator transition}

\author{T.\,I. Baturina}
\affiliation{Institute of Semiconductor Physics, 630090, Novosibirsk, Russia} 
\author{A.\,Yu. Mironov}
\affiliation{Institute of Semiconductor Physics, 630090, Novosibirsk, Russia} 
\author{V.\,M. Vinokur}
\affiliation{Materials Science Division, Argonne National Laboratory,
Argonne, IL 60439, USA}
\author{M.\,R. Baklanov}
\affiliation{IMEC Kapeldreef 75, B-3001 Leuven, Belgium}
\author{C. Strunk}
\affiliation{Institut f\"{u}r experimentelle und
angewandte Physik, Universit\"{a}t Regensburg, D-93025 Regensburg, Germany}

\date{\today}

\begin{abstract}
We investigate the insulating phase that forms in a titanium nitride film
in a close vicinity of the disorder-driven superconductor-insulator transition.
In zero magnetic field the temperature dependence of the resistance
reveals a sequence of distinct regimes upon decreasing temperature
crossing over from logarithmic to activated behavior with the variable-range
hopping squeezing in between.
In perpendicular magnetic fields below 2\,T, the thermally activated regime
retains at intermediate temperatures, whereas
at ultralow temperatures, the resistance increases
faster than that of the thermally activated type. 
This indicates a change of the mechanism of the conductivity. 
We find that at higher magnetic fields the
thermally activated behavior disappears
and the magnetoresistive isotherms saturate towards the value close to
quantum resistance $h/e^2$.
\end{abstract}


\maketitle

The subject of suppression of superconductivity by disorder in thin
superconducting films can be traced back to a pioneering work of early
forties by Shal'nikov~\cite{Shalnikov40}, where it was noticed for the first time,
that the superconducting transition temperature, $T_c$, decreases with the decrease
of the film thickness.
Later, a number of experimental works revealed a drastic suppression
in $T_c$ in thin films with the growth
of sheet resistance~\cite{Strongin70,Raffy83,Graybeal84},
which is determined by either film composition or thickness and
serves as a parameter measuring the degree of disorder.
The observed behavior was in a good quantitative accord with
theoretical predictions by Maekawa and Fukuyama~\cite{Maekawa}
and the subsequent work by Finkel'stein~\cite{Finkelstein}.
The physical picture behind suppressing superconductivity by disorder
is that in quasi-two-dimensional systems disorder inhibits
electron mobility and thus impairs dynamic screening of the
Coulomb interaction.
This implies turning on the Coulomb
repulsion between electrons which opposes Cooper attraction and,
if strong enough, breaks down Cooper pairing and destroys superconductivity.
Importantly, according to mechanism of~\cite{Finkelstein},
the degree of disorder at which the Coulomb repulsion would balance
the Cooper pair coupling is not sufficient to localize normal carriers;
thus at the suppression point superconductors transforms into a metal.
The latter can be turned into an insulator upon further increase of disorder.
Therefore this mechanism, which is referred to as a fermionic mechanism,
results in a sequential superconductor-metal-insulator transition~\cite{Finkelstein}.
There exists a seemingly alternative picture of the transition, the so-called
bosonic mechanism, where the intermediate metallic phase collapses
to a single point, implying the \textit{direct} disorder-driven
superconductor-to-insulator transition (D-SIT)~\cite{Gold,MPAFisher}.
Contrary to fermionic mechanism, the bosonic one realizes
via localization of the Cooper pairs, which survive even at the
nonsuperconducting side of the transition.
However, numerical simulations by Ghosal, Randeria, and Trivedi~\cite{Ghosal}
demonstrated that this distinction is not valid: near the D-SIT
the homogeneously disordered film breaks up into superconducting islands
separated by an insulating sea. Depending on the competition between
the charging energy of a single island and the Josephson coupling between
the neighboring islands, the film may become either superconducting 
or insulating, with the island-like structure maintaining 
at both sides of the transition.
Yet the details of
the microscopic mechanism of the D-SIT are far from being understood, and
uncovering the nature of the phase resulting from
suppression of the superconductivity by disorder remains
one of the major challenges of condensed matter physics.

In this work we focus on the insulating side of the D-SIT.
To begin with, we briefly summarize the up-to-date experimental
findings on the subject.
In the early experiments~\cite{Strongin70} with the very thinnest Pb films
it was noticed that once on the nonsuperconducting side,
the temperature dependence of resistance acquired an activated character:
\vspace{-3mm}
\begin{equation}
R=R_0\exp(T_0/T)\,.
\label{activation}
\end{equation}
Study of the Bi films~\cite{Haviland, Liu} revealed three major regimes
of the temperature resistance behavior in the insulator domain depending
on the degree of disorder.
The films closest to D-SIT exhibit
comparatively weak insulating trend with conductance lowering
as $\log T$ upon decreasing temperature. Moderately disordered films,
that are more far from the D-SIT, demonstrate the Efros-Shklovskii (ES) behavior,
$R=R_1\exp[(T_{ES}/T)^{1/2}]$,
and the resistance in most disordered films is thermally activated
(\ref{activation}).
The direct D-SIT was found in InO$_x$ films~\cite{ShaharOvadyahu},
exhibiting a sequence of temperature behaviors on the nonsuperconducting side:
deep in the insulating regime, the conductivity shows
Mott's variable range hopping (VRH)~\cite{ShaharOvadyahu,KowalOvadyahu}, on approach
to the transition with decreasing disorder it changes to ES law~\cite{KowalOvadyahu},
and the films closest to D-SIT shows activation (\ref{activation}),
which transforms to Mott's VRH at larger $T$~\cite{KowalOvadyahu}.
Interestingly, the crossover from the activation to ES law, and eventually
to Mott's VRH regime was also observed in InO$_x$ composite films~\cite{Kim}
upon increasing temperature.
It was noticed in Ref.~\cite{ShaharOvadyahu} that
applying a relatively small perpendicular magnetic field (0.7\,T)
to the least disordered insulating films results in positive magnetoresistance,
while more resistive samples show purely negative magnetoresistance.
Gantmakher's group carried out measurements of the effect of
the magnetic field up to 20\,T on the insulating InO$_x$ films in
the vicinity of the D-SIT~\cite{VFGInOIns}
and revealed nonmonotonic magnetoresistance behavior: a
positive magnetoresistance at low magnetic fields
turning into a negative magnetoresistance upon
increasing the field.
Importantly, several works, starting from the
earliest~\cite{ShaharOvadyahu,VFGInOIns} and ending by recent~\cite{Shahar-act}
demonstrate that in the zero magnetic field the resistance of insulating films
closest to the transition deviates \textit{downward} from
the Arrhenius activated behavior at lowest temperatures.
At high magnetic fields, the charge transfer mechanism appears
the same as that at the zero field but high temperatures and follows
 Mott's VRH law~\cite{VFGInOIns}.
Another recent finding worth noticing is the dependence
of the activation energy in InO$_x$ films on the linear
size of the film (the distance between the leads)~\cite{Ovadyahu}.
The D-SIT in TiN was found for the first time on the films prepared
by magnetron sputtering~\cite{Hadacek}.
On the insulating side close to the transition,
the resistance exhibited thermally activated
behavior till lowest temperatures available.
In our preceding works~\cite{BaturinaIns,BaturinaPhysC},
where we were dealing with the thin TiN films grown by atomic layer chemical
vapor deposition, an exceptionally sharp D-SIT was found.
At zero and low magnetic fields the resistance displayed thermally
activated behavior with the nonmonotonic dependence
of activation temperature on the magnetic field.
Here we present the novel results on identically prepared samples
showing that at the lowest temperatures the resistance deviates \textit{upward}
from the Arrhenius activated dependence (1)
in a contrast to observation in InO$_x$
films~\cite{ShaharOvadyahu,VFGInOIns,Shahar-act}.
Hereafter we will referring to this behavior as to \textit{hyperactivation}.
This behavior holds at low magnetic fields.
At higher magnetic fields, instead of this low-temperature upturn,
the resistance shows \textit{downward} deviation from the Arrhenius activation.
The analysis of the corresponding magnetoresistance shows
that it decays exponentially with field towards
the finite value close to the quantum resistance $R_\mathrm{Q}=h/e^2$.

The resistance was measured by the four-terminal technique in the linear
$I$-$V$ regime even at very low temperatures.
The width of the film was 50\,$\mu$m, the distance between the voltage probes
was 100\,$\mu$m, thus the sample comprised two squares.
All the data are presented as resistance per square.
The samples were cooled down in the Oxford 200 TLE dilution refrigerator.
Magnetic fields up to 10\,T were applied perpendicular to the film surface.

\begin{figure}[b]
\centerline{\includegraphics[width=70mm]{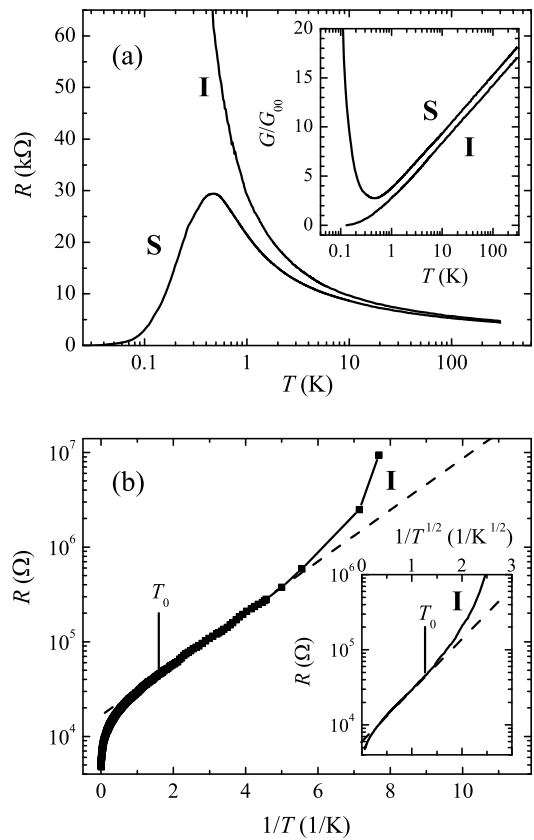}}
\caption{Zero magnetic field temperature behavior of the
resistances of samples \textbf{S} and \textbf{I}.
Panel\,(a): $R$ versus $\log T$ plots for \textbf{S} and \textbf{I}
samples are nearly undistinguishable at high temperatures.
The inset: the same data replotted as $G/G_{00}=\pi h/(e^2R)$ versus $\log T$
show almost identical logarithmic temperature dependencies of both samples
in the range 6 -- 300\,K.
Panel\,(b): $\log R$ versus $1/T$ plot for the \textbf{I}-sample displays
Arrhenius activation, the activation temperature being $T_0=0.63$\,K, determined
by fit to Eq. (1) the data in the temperature range 0.25 -- 0.9\,K.
The line bar marks this $T_0$.
At lowest temperatures $\log R(T)$ turns upwards
departing from the Arrhenius- to \textit{hyperactivated} regime.
The inset: the same data plotted as $\log R$ vs. $1/T^{1/2}$, demonstrate
the ES regime wedging in between the logarithmic and
activated transport. The fitting parameters are: 
$T_{ES}=2.6$\,K and $R_1=6.2$\,k$\Omega$.
}
\label{fig1}
\end{figure}
We start with the zero magnetic-field results obtained during the cooling
the cryostat down from room temperature.
Shown in Fig.\,1a are the temperature dependences of the resistance $R(T)$
of two samples, \textbf{S} and \textbf{I}.  
At room temperature, their resistances are close
($R_{300}=4.48$\,k$\Omega$ and $R_{300}=4.76$\,k$\Omega$, respectively), 
but diverge upon decreasing temperature. 
Namely, the \textbf{S}-sample
falls into a superconducting state, whereas the \textbf{I}-sample
becomes an insulator.
Upon cooling down to 3\,K both samples exhibit
nearly identical logarithmic temperature dependence of the conductance
(see inset in Fig.\,1a), which is well described by the formula
$G(T)/G_{00}=A\ln(k_BT\tau/\hbar)$, where $G(T)=1/R(T)$
is the conductance, $G_{00}=e^2/(\pi h)$, and $A=2.55$ for both samples.
This behavior is in accord with the theory of quantum corrections
for quasi-two-dimensional disordered systems and can be attributed
to localization and repulsive electron-electron interaction
corrections~\cite{AAreview}.
To characterize the behavior of the \textbf{I}-sample
at low temperatures we replot $R(T)$ versus $1/T$ in Fig.\,1b.
In the temperature interval between 0.25 and 0.9\,K,
the resistance is well fitted by a thermally activated dependence
of Eq.\,(1), with $T_0=0.63$\,K being the activation temperature,
and $R_0\approx 17$\,k$\Omega$.  Plotting the same data as $\log R$ versus
$1/T^{1/2}$ one can see that the resistance can be nicely fitted by the
ES-law in the higher temperature range 0.6 -- 6\,K (see inset in Fig.\,1b).
Note, that the deviation from the ES-law towards the activation starts
at the temperature $T\lesssim T_0$.
To summarize here, the evolution of $R(T)$ of the
insulating TiN film in the zero magnetic field is as follows.
As temperature decreases, the resistance
shows first a logarithmic growth crossing over into the ES-law, which
in its turn, transforms into the Arrhenius activated behavior, with all three
subsequent regimes pairwise overlapping.
So far the temperature behavior of $R(T)$ in TiN films more or less
parallels that of InO$_x$~\cite{Kim,KowalOvadyahu}.
An important difference arises at the very lowest
temperatures where $R(T)$ of TiN shoots up from the Arrhenius dependence
contrasting the downturn observed
in InO$_x$~\cite{ShaharOvadyahu,VFGInOIns,Shahar-act}.

Turning on the magnetic field we find its huge effect on the transport
properties of the insulating TiN films (see Fig.\,2).
Even a small ramping of
the field from zero to 0.1\,T at 220\,mK causes at least the 30 times
increase in the resistance; more specifically, the resistance
grows from 2.9$\cdot 10^5\,\Omega$ to $10^7\,\Omega$, which is
the upper limit of the used measuring circuit.
The magnetoresistance demonstrates pronounced nonmonotonic behavior:
an abrupt raise and rapid drop of the resistance
in the window of about 1.5\,T wide, the latter transforming into a
flattening tail at large fields.
At $T=220$\,mK, the negative magnetoresistance of the insulating sample
exceeds two orders of magnitude, and at 10\,T resistance drops
down to $R=46$\,k$\Omega$.
Another interesting feature of the magnetoresistance is a gradual shift of the
peak to higher fields upon increasing the temperature.
This behavior is clearly seen in the inset showing enlarged images
of the subsequent three highest temperature isotherms.
\begin{figure}
\centerline{\includegraphics[width=70mm]{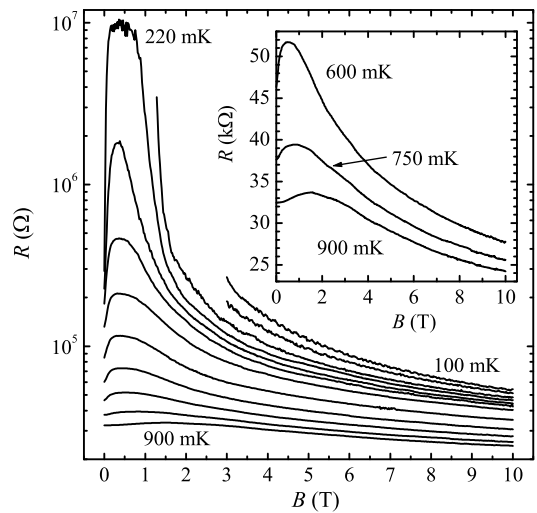}}
\caption{Magnetoresistance isotherms for the sample \textbf{I}
taken at temperatures
$T=100$, 140, 190, 220, 240, 260, 300, 380, 480, 600, 750, 900\,mK
respectively from the top to the bottom.
The inset shows a close-up view of the $R(B)$ curves measured at
$T=600$, 750, 900\,mK.}
\label{fig2}
\end{figure}
\begin{figure}
\centerline{\includegraphics[width=70mm]{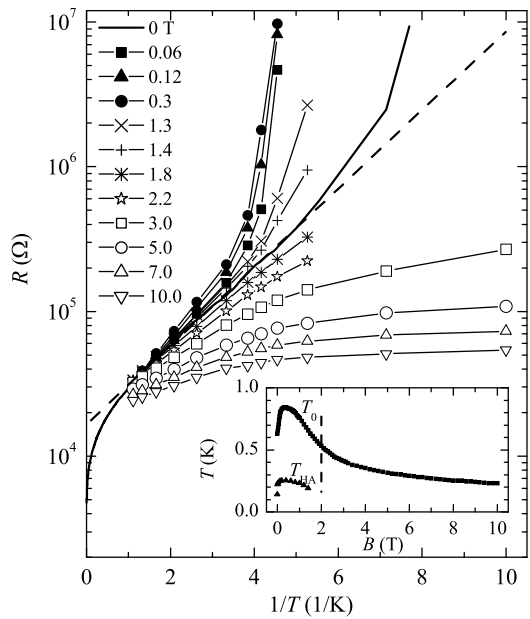}}
\caption {Arrhenius plots of the isomagnetic temperature dependencies
of the resistance of the insulating sample.
The inset: $T_0(B)$, the activation temperature extracted from $R(T,B_i)$
data in the temperature range 0.3 -- 0.9\,K. It assumes the maximum
value 0.84\,K at $B=0.32$\,T.
$T_\mathrm{HA}(B)$ determined as explained in the text.
}
\label{fig3}
\end{figure}
\begin{figure}[b]
\centerline{\includegraphics[width=75mm]{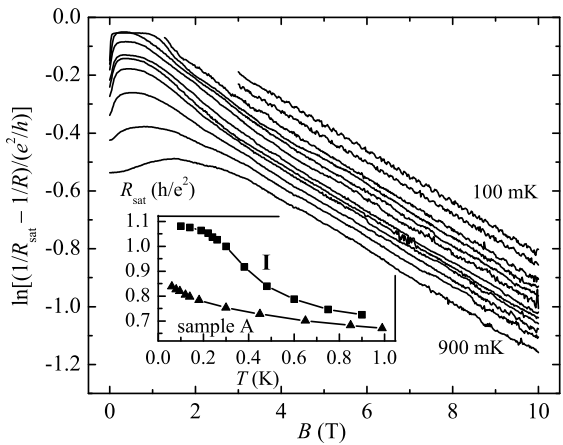}}
\caption{Scaling plot of the data shown in
Fig.\,2: For certain values of $R_\mathrm{sat}$,
$\ln\left[1/R_\mathrm{sat}-1/R(B)\right]$ varies linearly
vs.~$B$, with a $T$-independent slope. The linear slope corresponds
to the characteristic field $B^*\simeq 11.2$~T.
Inset: Temperature dependence of
$R_\mathrm{sat}$ for \textbf{I}-sample.
We added data for superconducting sample A from Ref.~\cite{TBPRLQM}.
}
\label{fig4}
\end{figure}

Taking the fixed-field cuts across the data presented in Fig.\,2, we
find that in the temperature range 0.3 -- 0.9\,K all the $\log R(T,B_i)$ 
versus $1/T$ plots are linear at all magnetic fields. 
Thus, making use of the Eq. (1),
we have determined field dependences $T_0(B)$ (shown in the inset of Fig.\,3)
and $R_0(B)$ (which vary between 13 and 21\,k$\Omega$).
The nonmonotonic behavior of $T_0(B)$ resembles that
of the magnetoresistance.
At very lowest temperatures, however, the $\log R(1/T)$ plots
deviate from the straight lines and diverge in a fan-like manner.
To make the observed features even more transparent we have grouped
the symbols for the data in Fig.\,3 into three branches.
The solid symbols correspond to the
temperature dependences taken within the \textit{positive} magnetoresistance
field window, i.e., at the magnetic fields $B_i< 0.32$\,T.
The `crosses' correspond to fields $B_i$ between 0.32\,T and 2\,T,
at the negative slope of the peak.
And, finally, open symbols denote isomagnetic temperature dependences
at high fields $B_i > 2$\,T, corresponding to the negative magnetoresistance tails.
The zero field resistance is shown as a solid line, and the slope
of the dashed line corresponds to the activation temperature at zero field.
Figure 3 shows that in relatively small magnetic fields the
resistance curves sharply jump upward from the Arrhenius
activation behavior below some magnetic field dependent temperature
$T_\mathrm{HA}(B)$.
This hyperactivated behavior maintains till fields not exceeding 2\,T.
We define the characteristic temperature of the transition to the
hyperactivated regime by the relation
$R(T_\mathrm{HA})=2R_0\exp(T_\mathrm{HA}/T_0)$, i.e.,
as the temperature at which the actually measured resistance
exceeded two times the resistance obtained by extrapolation from the
activated temperature interval.
The inset in Fig.\,3 shows magnetic
field dependence of $T_\mathrm{HA}(B)$ which appears
nonmonotonic and looks similar to the field dependence of $T_0$.
It is noteworthy that the hyperactivated regime exists on the both,
positive- (solid symbols) and negative (crosses)
magnetoresistance sides of the peak.
Upon the further increasing magnetic field,
the hyperactivation vanishes and is replaced
by the \textit{flattening} or downward deviation from the presumed
activated behavior (open symbols).
We emphasize here that the linearity of the $\log R(1/T)$ dependence
in the 0.3 -- 0.9\,K window at fields beyond 3\,T, can by no means be
taken as an evidence of the thermally activated carrier transfer, since
the activation temperatures that come out from fitting procedure
are very close or even less than the lower bound of the
observation interval, $T_0 \lesssim 0.3$\,K.
Moreover, the data in the whole temperature range $T<0.9$\,K and fields
beyond 3\,T are much better linearized in the $\log R$
versus $1/T^{1/2}$ representation.  
This suggests a recovery of the ES charge transfer mechanism
at intermediate fields, which for the zero magnetic
field was observed at $T>0.6$\,K.
We find that upon increasing the magnetic field the ES characteristic
temperature drops, whereas the preexponential factor increases, in particular, 
at $B=3$\,T, $T_{ES}=1.8$\,K and $R_1=6.8$\,k$\Omega$, and
at $B=10$\,T, $T_{ES}=0.39$\,K and $R_1=12.6$\,k$\Omega$.  
This indicates that the ES VRH is suppressed by high magnetic fields 
(judging from decreasing $T_{ES}$)
and that in the domain of the negative magnetoresistance the electron transfer
which at moderated magnetic fields is mediated by Cooper pairs hopping,
becomes quasiparticle dominated process at the elevated fields.
We expect that 
upon further rise of the magnetic field the ES behavior changes
to a metallic-like transport.  
At the same time, the behavior of the isotherms $R(B)$ of the insulating
sample at high $B$ resembles that of the superconducting samples from
the Ref.~\cite{TBPRLQM}, where it was described by the empirical relation
$G(T,B) =1/R_\mathrm{sat}(T)- \beta(T)\exp(-B/B^*)$,
with $B^*$ being some characteristic field, and
with $R_\mathrm{sat}$ being close to the quantum resistance $h/e^2$.
Following the same treatment as in~\cite{TBPRLQM},
we present the $R(B)$ data in a similar scaling form as
$G(B) =1/R_\mathrm{sat}- 1/R(B)$ and plot it on a logarithmic
scale as function of $B$ (see Fig.\,4).
Indeed, all the isotherms measured in the window from 900\,mK down to
100\,mK are well linearized via the proper choice of a single fitting
parameter $R_\mathrm{sat}$ for each temperature (shown in the inset in Fig.\,4).
The scaled $1/R(B)$ have the same temperature independent slope,
corresponding to the characteristic field $B^*\simeq 11.2$\,T,
which is close to $B^*\simeq 10.7$\,T for the sample A from Ref.~\cite{TBPRLQM}.

We would like to complement our findings with the data on low temperature
scanning tunneling spectroscopy measurements of the local density of states
in the identically prepared TiN films~\cite{STM_TiN}.
It was found that even in comparatively less weakly disordered films,
judging from the room temperature resistance,
the superconducting state appears inhomogeneous.
Furthermore, in spite of the fact, that the suppression of the superconducting
transition temperature follows the fermionic mechanism~\cite{Finkelstein}, 
the analysis of the whole aggregate of the data supports the idea that the
film in the critical region of the D-SIT can be viewed as
an array of the superconducting islands immersed into an insulating
sea or self-organized two-dimensional Josephson junction
array~\cite{Ghosal,KowalOvadyahu,VFGInOIns,BaturinaIns,BaturinaPhysC,Imry}.

To gain an insight into the hyperactivation behavior from this standpoint,
we juxtapose our data with the corresponding findings on granular 
films~\cite{Dynes78} and on two-dimensional Josephson junction arrays 
(2DJJ)~\cite{Tighe,Haviland2D,Mooij,Kanda,Yamaguchi}.
These systems showed two distinct activated regimes with the 
different Arrhenius slopes.  
The smaller one is observed in 
the high fields and/or high temperature domain,
where the superconductivity in the granules/superconducting electrodes was
suppressed.
In the low-$T$/low-$B$ region,
where granules/superconducting electrodes became superconducting,
the Arrhenius activated curves acquired the \textit{larger} slope.
Specifically, in 2DJJ arrays, the activation energy in the ``normal" state,
which was found to be close to $(1/4)E_c$, where $E_c$ was the charging energy 
of a single junction, increased to  $(1/4)E_c+\Delta_0$, with $\Delta_0$ 
being the gap of a superconducting electrode~\cite{Tighe,Haviland2D,Mooij}.
However, having decided to straightforwardly ascribe
the upturn in $\log R$ versus $1/T$ dependencies measured in our TiN
films to opening the superconducting gap, we would have encountered 
immediate problems.  Namely, using, for example, our data
at $B=0.06$\,T, where the low temperature activation temperature
as crudely estimated from the upturn is $T_0^l\simeq 6$\,K, while at high
temperatures $T_0=0.71$\,K, we arrive at the magnitude of the opening gap
to be of about 0.6 of that of the bulk TiN~\cite{Escoffier}. 
This is in strong conflict with the STM data of Ref.~\cite{STM_TiN}, 
where this ratio is estimated not to exceed 0.1 at the insulating
side near the D-SIT.
Furthermore, having assumed that the observed upturn reflects the activation
behavior, i.e. switching from a low activation energy to higher one
we derive from the above analysis the nonphysically small pre-exponential 
factor, $R_0^l\approx 1.3\cdot 10^{-5}\,\Omega$.  
This makes the notion
of the activation transport at lowest temperatures in our films meaningless.
On the other hand, Kanda and Kobayashi, and later Yamaguchi 
\textit{et al}~\cite{Kanda,Yamaguchi},
observed that the resistance of the 2DJJ array in the regime where electrodes 
are superconducting, increases, with decreasing 
temperature, faster than that of the thermal-activation type.  
They interpreted it as a signature of the charge binding-unbinding 
Berezinskii-Kosterlitz-Thouless (C-BKT) transition~\cite{Fazio}.
We propose that the hyperactivated behavior observed in our experiments
is of the same nature and signalizes the formation of a low temperature 
C-BKT-phase, which we call a
\textit{superinsulator}~\cite{FVB,VinNature}.

In conclusion, we have investigated the temperature- and the magnetic 
field dependencies of the resistance of thin TiN films in the critical 
region of the disorder-driven superconductor-insulator transition.  
We have found that quantum metallicity, i.e.
the saturation of high field negative magnetoresistance near the $h/e^2$ 
is a generic property of these films which belong not only to the superconducting-
but also to the insulating side of the transition. At the same time, 
in the low perpendicular magnetic fields domain the thermally activated 
regime takes place at intermediate temperatures. At ultralow
temperatures, the resistance becomes hyperactivated, i.e., grows faster than that 
of the thermally activated type.  We relate the change of the mechanism of 
the conductivity with the charge binding-unbinding Berezinskii-Kosterlitz-Thouless
transition and refer to the appearing low-temperature BKT-phase as to a
superinsulating state.  The conditions of formation of a
superinsulating phase accounting for the role of the finite-size 
effects and the finite-range charge screening require further research.

\begin{acknowledgments}
We are grateful to A. Gerber for fruitful discussions.
This research has been supported by the
RFBR Grant No. 06-02-16704, the U.S. Department of Energy Office of
Science through contract No. DE-AC02-06CH11357, 
and the Deutsche Forschungsgemeinschaft within the GRK 638.
\end{acknowledgments}

\end{document}